\documentstyle[12pt,epsfig,amsmath,amssymb]{article}
\begin{document}

\noindent September 2009\\

\vspace{5mm}

\begin{center}

{\Large PENTAGRAMS AND PARADOXES}

\vspace{18mm}

\hspace{5mm} Piotr Badzi{\c a}g*$^1$ \hspace{10mm} 

\vspace{8mm}

Ingemar Bengtsson*$^2$
\hspace{30mm} Ad\'an Cabello**$^3$

\vspace{8mm}
 
Helena Granstr\"om***$^4$
\hspace{10mm} Jan-\AA ke Larsson****$^5$

\vspace{18mm}

*{\it Stockholms Universitet, AlbaNova, Fysikum, 106 91 Stockholm}

**{\it Dep. F\'isica Aplicada II, Universidad de Sevilla, 410 12 Sevilla}

***{\it Matematiska Institutionen, Stockholms Universitet, 106 91 Stockholm}

****{\it Matematiska Institutionen, Link\"opings Universitet, 581 83 Link\"oping}

\vspace{10mm}

{\bf Abstract}

\end{center}

\vspace{3mm}

\noindent Klyachko and coworkers consider an orthogonality graph in the form 
of a pentagram, and in this way derive a Kochen-Specker inequality for spin 
1 systems. In some low-dimensional situations Hilbert spaces are naturally 
organised, by a magical choice of basis, into $SO(N)$ orbits. Combining these 
ideas some very elegant results emerge. We give a careful discussion of the 
pentagram operator, and then show how the pentagram underlies a number of 
other quantum ``paradoxes'', such as that of Hardy.

\vspace{8mm}

\begin{center}

{\tiny $^1$pbg01@physto.se \ \ $^2$ingemar@physto.se \ \ $^3$adan@us.es \ \ 
$^4$helenag@math.su.se \ \ $^5$ jalar@mai.liu.se}

\end{center}

\newpage

{\bf 1. Introduction}

\vspace{5mm}

\noindent We will be interested in five unit vectors. Counting them modulo 
5 we assume the orthogonality relations 

\begin{equation} \langle k|k+2\rangle = 0 \ , \hspace{8mm} k \in \{ 0,1,2,3,4\} 
\ . \label{ett} \end{equation} 

\begin{figure}[h]
        \centerline{ \hbox{
                \epsfig{figure=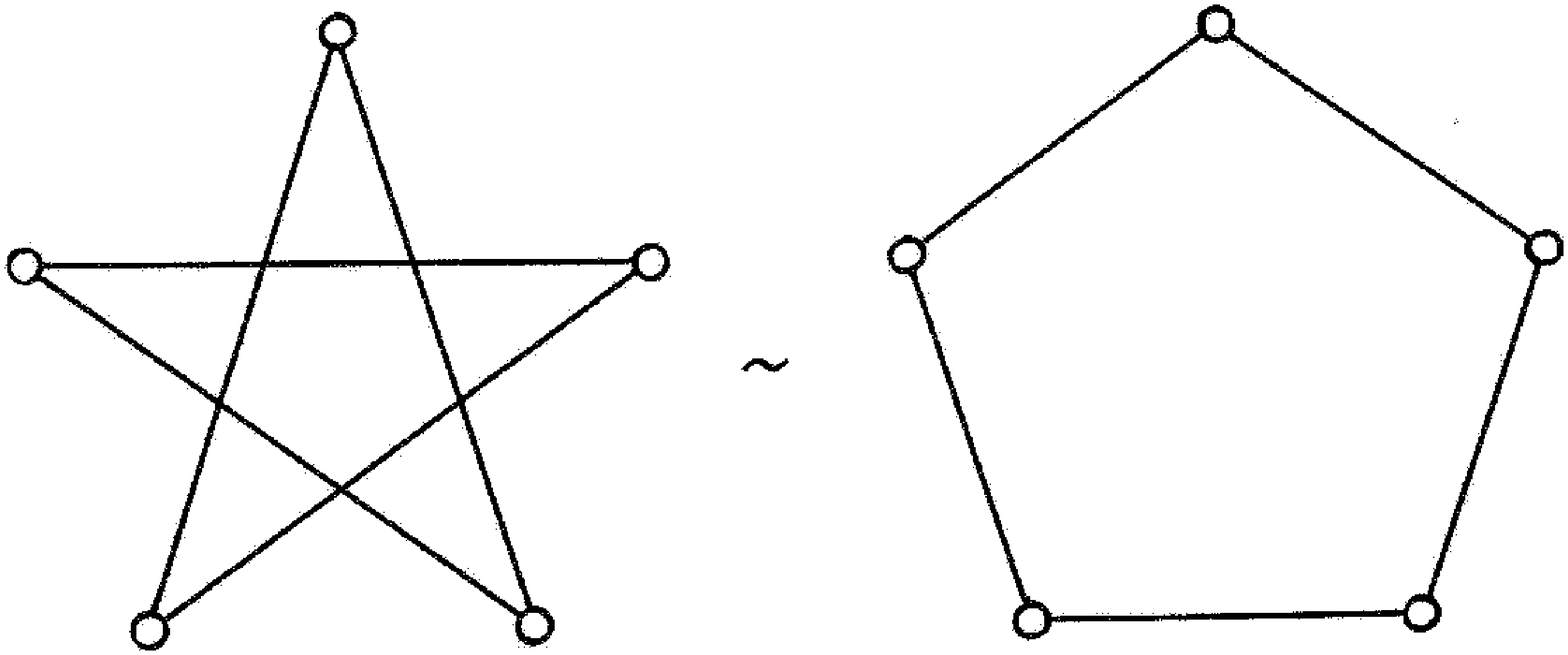,width=100mm}}}
        \caption{\small Two equivalent ways to draw our orthogonality graph. Since 
orthogonal vectors are far apart the pentagram is arguably a more 
``realistic'' picture.}
        \label{fig:pentagram0}
\end{figure}

\noindent To visualise this we draw an orthogonality graph, where each 
node represents a vector and each link an ortogonality relation. Our 
orthogonality graph is therefore a pentagram---or a pentagon, depending 
on how we draw it (see Fig. \ref{fig:pentagram0}). The pentagram operator is  

\begin{equation} \Sigma \equiv \sum_{i=0}^4 |k\rangle \langle k| \ . 
\label{tva} \end{equation}

\noindent The dimension of Hilbert space is assumed to be 3 or 4, so 
that the pentagram vectors form an overcomplete set, and in fact a frame 
of a peculiar kind.    

In 3 dimensions any pair of vectors uniquely determines a third ray, a 
vector up to phase. In this way a non-degenerate pentagram uniquely determines 
5 orthogonal triads. These were used by Wright in an interesting contribution 
to quantum logic \cite{Wright, Svozil}. He explored probability assignments 
consistent with Gleason's rules, and found that for the pentagram there 
exists assignments that are inconsistent with quantum mechanics when the 
graph is embedded in a full Hilbert space. 

This is already enough to show that something interesting can be done with 
pentagrams. More recently Klyachko and coworkers \cite{Can, Klyachko} 
derived a pentagram inequality that---following Kochen and Specker 
\cite{KS}---shows quantum mechanics to be probabilistically 
inconsistent with an underlying non-contextual reality. Here is their argument 
in a slightly modified form \cite{Bub}: The Kochen-Specker rules demand that 
each node in the graph is assigned a value $0$ or $1$ (a preassigned truth 
value), in such a way that orthogonal vectors are never both assigned the value 
$1$, and such that the sum of the values in any orthogonal basis equals $1$. 
When the same assignments are carried over to 
the projectors in the pentagram operator we see that at most two of them 
can be assigned the value $1$. In a non-contextual reality an experimenter, 
when adding observed frequencies of the five projectors she can measure, 
will therefore always find that 

\begin{equation} \langle \Sigma \rangle \leq 2 \ , \label{tre} \end{equation}

\noindent regardless of how she prepares her state. The point of course 
is that the quantum expectation values of $\Sigma$ will violate this bound 
for appropriately chosen states.

In section 2 we give a thorough discussion of the three dimensional 
pentagram operator. In section 3 we impose additional restrictions on 
the frame vectors. Thus we can insist that they all share the same degree 
of entanglement, or spin coherence. If we use the magical basis, available 
in 3 and 4 dimensions, maximally entangled or non-coherent states are real 
vectors, and conversely \cite{Hill, Uhlmann, Klyachko}. We can always form 
a pentagram from real vectors. In 3 dimensions we then find the beautiful 
feature that every non-coherent state will violate some pentagram inequality 
\cite{Klyachko}---in complete analogy with the fact that every entangled 
state in four dimensions violates some Bell inequality. In sections 4 and 5 
we go on to show that the pentagram inequality unifies several 
quantum ``paradoxes'' with slightly more involved orthogonality graphs. 
This includes a graph used by Kochen and Specker themselves \cite{KC}, the 
Aharon-Vaidman game \cite{Vaidman}, and Hardy's paradox \cite{Hardy}. In 
section 6 we briefly discuss the four dimensional pentagram operator, 
and finally suggest some open questions that we find interesting.

\vspace{3cm}

\vspace{1cm}

{\bf 2. The pentagram operator}

\vspace{5mm}

\noindent The largest violation of the KS inequality (\ref{tre}) will always 
be achieved by the eigenvector corresponding to the largest eigenvalue, so 
with a view to understand what violations are possible we first discuss 
the possible spectra of the pentagram operator $\Sigma$. A quick way to 
see what spectra can occur is the following. Define 

\begin{equation} p_{k,k+1} = |\langle k|k+1\rangle |^2 \ . \end{equation}

\noindent A calculation shows that 

\begin{equation} \mbox{Tr}\Sigma = 5 \hspace{12mm} \mbox{Tr}\Sigma^2 = 5 + 2A 
\hspace{12mm} \mbox{Tr}\Sigma^3 = 5 + 6A \ , \end{equation}

\noindent where 

\begin{equation} A = \sum_{k=0}^4p_{k,k+1} \ . \end{equation}

\noindent Assuming that the dimension of Hilbert space is three, the eigenvalues 
of $\Sigma$ therefore obey the characteristic equation  

\begin{equation} P_3(\lambda ) = \lambda^3 - 5\lambda^2 + (10-A)\lambda + 3A - 10  
= 0 \ . \label{kar3} \end{equation}

\noindent From this we see that there is a one parameter family of possible spectra 
labelled by the quantity $A$. So we need to know all possible values of $A$.

Let us therefore write down the most general set of five unit vectors obeying 
(\ref{ett}), and then compute $A$ from the corresponding pentagram operator. 
We adapt a basis to the orthogonal vectors $|0\rangle $ and $|3\rangle $. Up to 
irrelevant phase factors we find the five column vectors 

\begin{equation} \begin{array}{cccccccccrr} 1 & \ & \cos{a} & \ & 0 & \ & 0 & \ & \ & 
e^{-i\mu}\sin{a}\cos{b}/\sqrt{1 - s_a^2s_b^2} \\
0 & \ & 0 & \ & \cos{b} & \ & 1 & \ & \ & e^{-i\nu}\cos{a}\sin{b}/\sqrt{1 - s_a^2s_b^2}  \\
0 & \ & e^{i\mu}\sin{a} & \ & e^{i\nu}\sin{b} & \ & 0 & \ & \ & 
- \cos{a}\cos{b}/\sqrt{1 - s_a^2s_b^2} & \ .  
\end{array}  \label{familj} \end{equation}

\noindent We see that the set of all pentagrams is labelled by four angular 
variables. We will distinguish the two parameter set of real pentagrams for 
which all the five vectors are real, the one parameter set of real symmetric 
pentagrams for which $a = b$, and the regular pentagram for which the 
five real vectors form a cone with a regular pentagram as its base \cite{Can}. 
The pentagram is said to be degenerate if two 
of its projectors coincide. 

The spectrum of an arbitrary pentagram operator is determined by 

\begin{equation} A = 2 - \frac{\sin^2{a}\sin^2{b}\cos^2{a}\cos^2{b}}
{1-\sin^2{a}\sin^2{b}} \ . \end{equation}

\noindent Evidently any allowed value of $A$ and hence any spectrum can 
be obtained from a real pentagram, and in fact from a real symmetric pentagram. 
It is straightforward to show that 

\begin{equation} A_{\rm min} \leq A \leq 2 \ . \end{equation}

\noindent The maximum is obtained by a degenerate pentagram. 
The minimum $A_{\rm min} \approx 1.91$ is attained by the regular pentagram 

\begin{equation} \sin^2{a} = \sin^2{b} = \Phi - 1 \hspace{5mm} \Rightarrow \hspace{5mm} 
A = A_{\rm min} = 2 - \frac{1}{\Phi^5} \ , \label{Amin} \end{equation}

\noindent where $\Phi$ is the Golden Mean 

\begin{equation} \Phi = \frac{1}{\Phi - 1} \ , \hspace{8mm} \Phi = \frac{1+\sqrt{5}}{2} 
\approx 1.618 \ . \end{equation}

\noindent We recall that the Golden Mean arises as the ratio between the sides 
of a triangle inscribed in a regular pentagon, with angles $72^\circ$, $72^\circ$, and 
$36^\circ$. The smaller triangle one of whose sides is obtained by taking the bisectrix 
of one of the larger angles is similar to the original triangle; this is how the 
equation for the Golden Mean arises. 

\begin{figure}[h]
        \centerline{ \hbox{
                \epsfig{figure=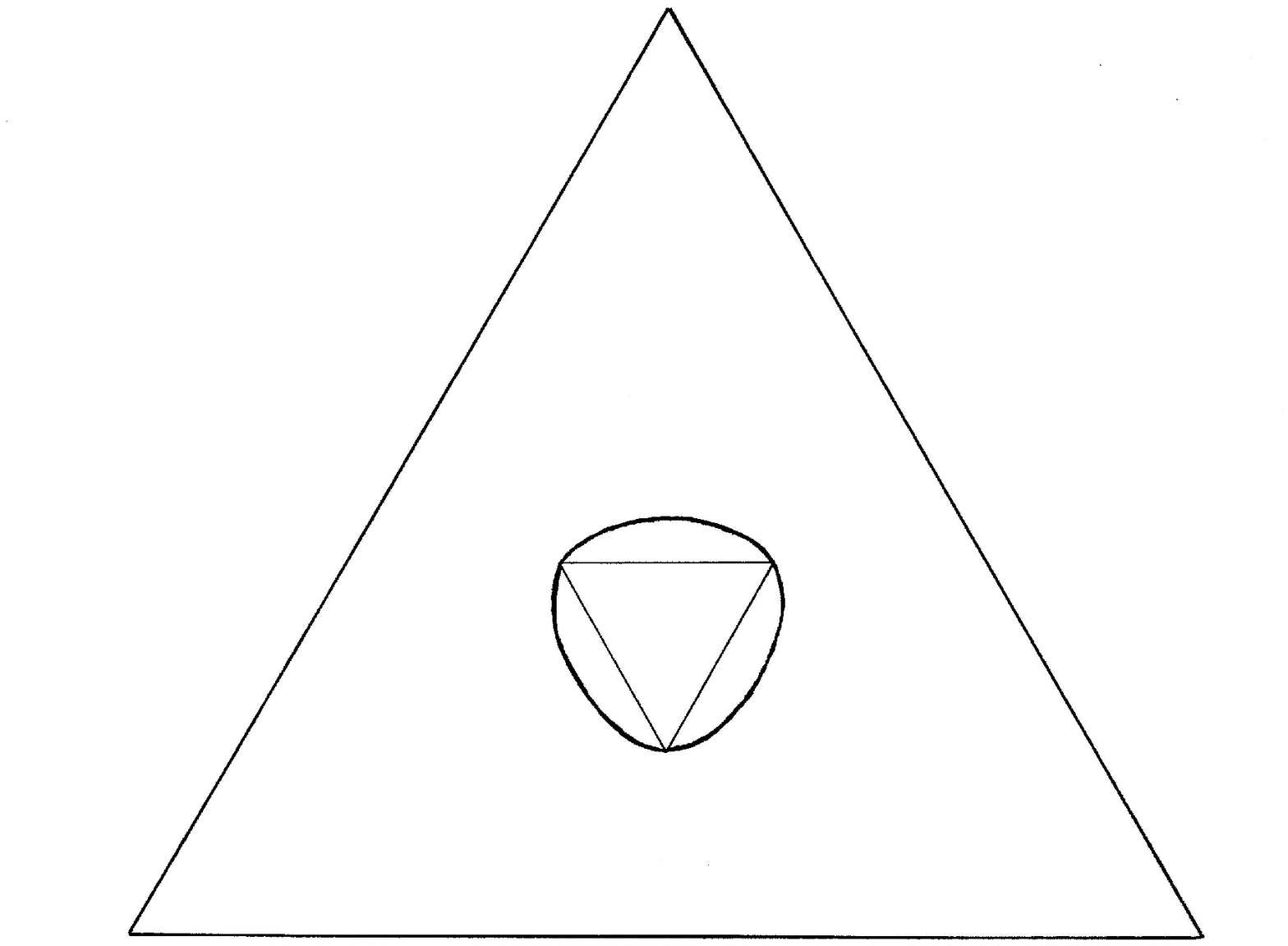,width=85mm}}}
        \caption{\small The spectra of all possible pentagrams as a curve in 
the eigenvalue simplex. The sum of the eigenvalues equals 5. At a corner 
they would be $(5,0,0)$. The inscribed triangle has a corner at $(2,2,1)$ and 
contains spectra that would not violate the KS inequality, should they occur 
(they do not, except at the corners).}
        \label{fig:pentagram1}
\end{figure}

The possible eigenvalues can now be computed explicitly as a function of the single 
variable $s = \sin{a}$, viz.

\begin{equation} \lambda_0 = 2-s^2 \ , \hspace{7mm} \lambda_\pm = \frac{3+s^2}{2} \pm \frac{1}{2}
\sqrt{\frac{1 + 3s^2 - 5s^4 + s^6}{1+s^2}} \ .
\end{equation}

\noindent This is perhaps not so illuminating, so  in Fig. \ref{fig:pentagram1} we 
display this one parameter family of spectra in an eigenvalue simplex. The largest 
possible eigenvalue is 

\begin{equation} \sqrt{5} \approx 2.236 \ . \end{equation}

\noindent The corresponding pentagram is regular. Real vectors then violate the 
inequality provided their angle with the eigenvector giving maximal violation is 
less than $31^\circ$; the pentagram vectors themselves give no violation. 

The smallest possible eigenvalue is $1$, and the spectrum $(2,2,1)$ results from 
degenerate pentagrams. These are the 
only ones that do not lead to any violation of the KS inequality (\ref{tre}). For 
later use we also observe that when $a = \epsilon$ is small the pentagram is 
almost degenerate and 

\begin{equation} \lambda_+ \approx 2 + \epsilon^2 \hspace{8mm} \lambda_0 \approx 
2 - \epsilon^2 \ . \label{epsilon} \end{equation}

\noindent These eigenvalues approach $2$ at the same rate. 

\vspace{1cm}

{\bf 3. The magical basis}

\vspace{5mm}

\noindent A vector in a complex Hilbert space can always be split into real 
and imaginary parts. Any vector can be written as 

\begin{equation} |\psi\rangle = \cos{\sigma}{\bf x} + i\sin{\sigma}{\bf y} 
\ . \label{1} \end{equation}

\noindent But a physical state is represented by an equivalence class 
$|\psi \rangle \sim e^{i\phi}|\psi \rangle$ of unit vectors, and every 
equivalence class contains a representative such that  

\begin{equation} {\bf x}^2 = {\bf y}^2 = 1 \ , \hspace{8mm} {\bf x}\cdot 
{\bf y} = 0 \ , \hspace{8mm} 0 \leq \sigma  \leq 
\frac{\pi}{4} \ . \label{2} \end{equation}

\noindent The phase is thereby fixed up to a sign (unless $\sigma = \pi/4$). 
This works in all complex Hilbert spaces, but is usually of no interest because 
unitary transformations will not preserve the scalar product ${\bf x}\cdot 
{\bf y}$. But in 3 dimensions, if we are interested in spin coherent 
states, then we have restricted ourselves to the subgroup $SU(2) \sim SO(3)$ 
of the full unitary group. This is an experimentally motivated restriction 
in some circumstances \cite{biphotons}. 
Similarly, if we are interested in pairs of entangled qubits we are restricted 
to the local subgroup $SU(2)\times SU(2) \sim SO(4)$. In both cases 
a magical basis is, by definition, a basis in which $SO(N)$ is represented by 
real matrices \cite{Hill}. Once a magical basis is adopted the parameter $\sigma$ in eq. 
(\ref{1}) is left unchanged by the transformations we do, and hence it labels 
the orbits of the relevant groups. We will refer to the quantity 

\begin{equation} C = |\langle \psi^\star|\psi \rangle | = \cos{2\sigma} 
\end{equation}

\noindent as the concurrence of the state. The 'classical' orbit 
with $C = 0$ consists of spin coherent states in 3 dimensions, and of 
separable states in 4. Real vectors have $C = 1$ and are 'maximally 
non-classical'; in 4 dimensions these are the maximally entangled states. 

There are various ways to substantiate the use of the adjective 'classical' 
here \cite{Klyachko, Kus}. A particularly relevant one, applicable in 3 
dimensions, is the following \cite{Klyachko}. First we restrict ourselves 
to pentagrams constructed from purely real vectors. The operator $\Sigma$ is 
then real, and can be diagonalised by transformations belonging to the real 
rotation group $SO(3)$. They leave the concurrence, and the 'classicality' 
of the states, unchanged. For a general state (\ref{1}) we tailor make a 
pentagram operator such that the eigenvector corresponding to the largest 
eigenvalue $\lambda_1$ is ${\bf x}$, and that corresponding 
to the second largest $\lambda_2$ is ${\bf y}$. Clearly 
 
\begin{equation} \langle \psi|\Sigma |\psi \rangle = {\bf x}\Sigma {\bf x} 
\cos^2{\sigma} + {\bf y}\Sigma {\bf y} \sin^2{\sigma} = 
\lambda_1\cos^2{\sigma} + \lambda_2\sin^2{\sigma} \ . \end{equation}

\noindent We now go back to Fig. \ref{fig:pentagram1} supplemented by eq. 
(\ref{epsilon}). Then we can convince ourselves that we obtain $\langle 
\Sigma \rangle = 2$ if and only if $\sigma = \pi/4$, that is for a coherent 
state, while all other states will violate its tailor made KS inequality. 
This is in complete analogy to the 
well known fact that all entangled states in 4 dimensions violate some 
Bell inequality \cite{who}.    

\vspace{1cm}

{\bf 4. Quantum paradoxes}

\vspace{5mm}

\begin{figure}[h]
        \centerline{ \hbox{
                \epsfig{figure=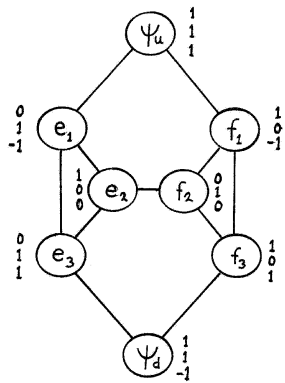,width=40mm}}}
        \caption{\small The Kochen-Specker subgraph. In 3 dimensions the upper 
pentagon completely determines the lower. An explicit realisation maximising 
the amount by which $|\psi_d\rangle $ violates the upper pentagram 
inequality is shown.}
        \label{fig:pentagram3}
\end{figure}

\noindent If we begin to examine all the Kochen-Specker type arguments that 
have been put forward in the literature we see that pentagons are ubiquitous 
in the orthogonality graphs that underlie them. Thus, consider the graph 
in Fig. \ref{fig:pentagram3}, which we refer to as the Kochen-Specker subgraph 
\cite{KS}. We assume that Hilbert space has 3 dimensions; then all the rays 
are uniquely defined once the five rays in the upper pentagon are given.  

The Kochen-Specker rules make it 
impossible to assign the value 1 to both the upper and the lower vector. 
In a non-contextual reality the system cannot have the properties 
corresponding to these two projectors at the same time. Nevertheless 
it is a realisable graph. With the choice of vectors shown in the figure 
we obtain 

\begin{equation} p = |\langle \psi_u|\psi_d\rangle |^2 = \frac{1}{9} \ . 
\label{probab} \end{equation}

\noindent This is a probabilistic violation of non-contextual reality \cite{Stairs}. 
It is also the maximal violation that one can obtain from this graph. Our point 
here is that maximising $p$ is equivalent to maximising the 
amount by which $|\psi_d\rangle$ violates the pentagram inequality for the 
upper pentagon in the graph. The proof is simple. As shown in Fig. the graph 
contains two ortogonal triads $|e_i\rangle_{i=1}^3$ and $|f_i\rangle_{i=1}^3$. 
Evidently 

\begin{equation} \langle \psi_d|e_1\rangle \langle e_1|\psi_d \rangle + \langle \psi_d | 
e_2\rangle \langle e_2|\psi_d \rangle = \langle \psi_d|\psi_d\rangle - 
\langle \psi_d|e_3\rangle \langle e_3|\psi_d \rangle = 1 \ , \end{equation}

\noindent and similarly for the other triad. It follows that 

\begin{equation} \langle \psi_d|\Sigma_u|\psi_d \rangle = \langle \psi_d|
\psi_u\rangle \langle \psi_u|\psi_d\rangle + 2 = p + 2 \ . \label{p+2} 
\end{equation} 
 
\noindent End of proof. 

The Kochen-Specker subgraph lies behind a number of quantum ``paradoxes''. 
Thus, consider the Aharon-Vaidman game \cite{Vaidman}, in which Alice prepares 
a particle. She can place it in one of two boxes, or in none of them, and 
then hands the boxes to Bob. Bob is allowed to open one box to 
see if it contains the particle, and then leaves it as he found it. 
At the end of each run Alice is allowed to perform a measurement and can decide 
to cancel that particular run of the game. For all the runs that she decides 
to count, she wins if Bob found the particle. (There is an on-looker who 
ensures that Bob does not cheat.) The question is: can Alice select the runs 
that are counted in such a way that she always wins? Classically the 
answer is clearly ``no'', in particular the extra option of not placing the 
particle in any of the boxes is worse than useless for her. But Alice has 
studied the Kochen-Specker subgraph with some care. She prepares the particle 
in the state 

\begin{equation} |\psi_u\rangle = |\mbox{first box}\rangle + 
|\mbox{second box}\rangle + |\mbox{no box}\rangle \ . \end{equation}

\noindent If Bob opens the first box and finds the particle, he leaves it 
in the state $|e_1\rangle = |\mbox{first box}\rangle$, if he does not find the particle 
in the first box he leaves it in the orthogonal state $|e_3\rangle = 
(|\mbox{second box}\rangle + 
|\mbox{no box}\rangle)$, and similarly if he opens the second box. At the 
end Alice makes a measurement corresponding to the projector 
$|\psi_d\rangle \langle \psi_d|$. If she gets the answer ``yes'' she knows 
that Bob found the particle. 

Since Alice knows that Bob did open the box one can argue that her state 
assignment has changed to an appropriate density matrix, but since she 
knows nothing about what Bob did or saw this does not change her probability 
to obtain ``yes''. This is still given by eq. (\ref{probab}), which means that she 
can select one ninths of all the runs in such a way that she always wins.

\vspace{10mm} 

{\bf 5. Hardy's paradox}

\vspace{5mm}

\noindent Another interesting application of the Kochen-Specker 
subgraph is to Hardy's paradox \cite{Hardy}, which has been translated into 
experiments \cite{DeM?}.  The paradox concerns four dichotomic variables, 
the pair $A_i$ controlled by Alice and the pair $B_i$ controlled by Bob. It is 
assumed that 

\begin{equation} P(B_2=0|A_1 = 1) = P(A_2 = 0|B_1 = 1) = P(A_2 = 1, B_2 = 1) = 0 
\ . \label{hardy} \end{equation}

\noindent Hence it would seem that if $A_1$ and $B_1$ were both true, then 
$A_2$ and $B_2$ would both have to be true too, but this never happens in a single 
experiment. A paradox therefore arises when 

\begin{equation} P(A_1=1, B_1 = 1) \neq 0 \ . \end{equation}

\begin{figure}[h]
        \centerline{ \hbox{
                \epsfig{figure=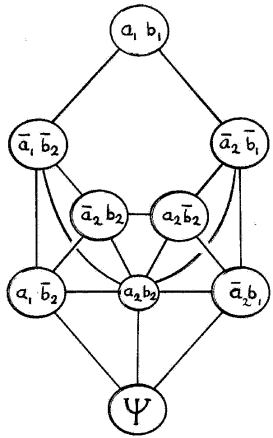,width=40mm}}}
        \caption{\small The orthogonality graph for Hardy's paradox; the notation 
is that ``$a_1\bar{b}_2$'' stands for the state where the eigenvalue of $A_1$ is $1$ 
and the eigenvalue of $B_2$ is $0$, etc. The orthogonalities enjoyed by the state 
$|\Psi\rangle $ ensure that eqs. (\ref{hardy}) hold.}
        \label{fig:pentagram2}
\end{figure}

\noindent But this logical structure is obtained by adding a single ``outlier'' 
to the Kochen-Specker subgraph; see Fig. \ref{fig:pentagram2} which also 
introduces some notation. All the states in the two pentagons except the 
upper vector $|a_1b_1\rangle $ are orthogonal to the outlier $|a_2b_2\rangle $. 
Therefore 7 of the 9 states in the graph lie in a three dimensional subspace. 
The state $|a_1b_1\rangle $ must have a component 
in this subspace so as to not coincide with the outlier, but it cannot be 
confined to it since $\langle a_1b_1|a_2b_2\rangle = 0$ would cause the 
construction to collapse.

This is very much a four dimensional paradox, since the entry of Alice and 
Bob requires the upper pentagram to be made of separable states only. The 
only vector in the graph which is not separable is $| \Psi \rangle$, the 
state that the system is actually in. But the outlier state $|a_2b_2\rangle $ 
ensures that all other states are uniquely determined by the upper pentagon, 
since all other states except $|a_1b_1\rangle $ are confined to three dimensions. 
As stressed by Penrose the logical structure is in a sense three dimensional 
\cite{Penrose}. 

The non-classical probability that we wish to maximise, under the constraint 
that all vectors in the graph except $|\Psi \rangle$ are separable, is 

\begin{equation} p = |\langle a_1b_1|\Psi \rangle |^2  \ . \end{equation}

\noindent The exact answer \cite{Mermin} is 

\begin{equation} p = \frac{1}{\Phi^5} \approx 0.09 \ . \end{equation}

\noindent 
Using the very same argument that led to eq. (\ref{p+2}) we can show that 
maximising $p$ is equivalent to maximising the amount by which $|\Psi \rangle $ 
violates the pentagram inequality for the upper pentagon in the graph.

\vspace{10mm}

{\bf 6. Unfinished work and open questions}

\vspace{5mm}

\noindent We did not complete the classification of all pentagrams in 4 
dimensions. We can choose the pentagram vectors to be either maximally 
entangled (real, with $C = 1$) or separable ($C = 0$). We can still 
define a regular pentagram as one in which the scalar products between the 
non-orthogonal vectors have the same moduli. Among separable pentagrams 
the regular pentagram is unique, and has the eigenvalues 

\begin{equation} \vec{\lambda} = (2.148, 1.470, 1.240, 0.142) \ . \end{equation}

\noindent The maximal violation of the KS inequality is somewhat less than the 
largest violation in 3 dimensions. For maximally entangled (real) pentagrams, 
there are two regular solutions. One of them is simply the three dimensional 
regular pentagram with an extra zero eigenvalue added, the other has 

\begin{equation} \vec{\lambda} = (1.809, 1.809, 0.691, 0.691) \end{equation}

\noindent and does not lead to a violation. We note that there exists a set of 
18 real vectors in 4 dimensions that are non-colourable, and lead to an absolute 
rather than probabilistic Kochen-Specker contradiction \cite{Cabello}. The 
corresponding orthogonality graph contains a pentagram subgraph \cite{Adan} 
whose eigenvalues turn out to be 

\begin{equation} \vec{\lambda} = (2.171, 1.235, 1.5, 0.093) \ , \end{equation}

\noindent leading to a rather large violation.     

The reader may wonder why we restrict ourselves to pentagons. Why not heptagons? 
In 3 dimensions the answer is simple: although there is a KS inequality associated 
to the heptagon, the maximal violation is smaller (in the precise sense that it is 
more sensitive to added noise \cite{IB}). It may be interesting to pursue heptagons 
in higher dimensions though.

Recently several experiments related to the Kochen-Specker theorem have been 
performed \cite{nagon, nagon2, Elias}. We would obviously be delighted to see an 
experimental violation of the pentagram inequality. The challenging part of 
such an experiment would be to verify that the Kochen-Specker rules apply in all 
cases where the theory says that they can be checked with simultaneous measurements. 

We end on a speculative note. Take any Kochen-Specker graph in 3 or 4 dimensions 
leading to an absolute rather than a probabilistic Kochen-Specker contradiction, 
such as the one briefly alluded to above. Enumerate all its pentagram subgraphs. 
Is it by any chance the case that every vector in Hilbert space will violate 
at least one of the corresponding pentagram inequalities?

\end{document}